\documentclass[aps, 11pt]{revtex4-1}

\usepackage{amssymb}
\usepackage{latexsym}
\usepackage{amsmath}
\usepackage{amsfonts}
\usepackage{amsopn}
\usepackage[latin1]{inputenc}
\usepackage{setspace}
\usepackage{color}
\usepackage{enumerate}
\usepackage{multirow}
\usepackage{microtype}
\usepackage{dcolumn}
\usepackage{bm,bbm}
\usepackage{upgreek}
\usepackage{graphicx}
\usepackage{overpic}
\usepackage{subfig}
\usepackage{natbib}
\usepackage{commath}
\usepackage{booktabs} 
\usepackage{tikz}
\usepackage[font=small]{caption}
\newcommand*{\Ca}{\text{Ca}}

\renewcommand*{\Re}{\operatorname{Re}}

\begin{document}
\singlespace

\title{The influence of van der Waals forces on a bubble moving in a tube}
\date{\today}

\author{Naima H. Hammoud}
\email{nhammoud@princeton.edu}
\affiliation{Program in Applied and Computational Mathematics, Princeton University, Princeton, NJ 08544, USA}
\author{Philippe H. Trinh}
\affiliation{Oxford Centre for Industrial and Applied Mathematics\vphantom{}, Mathematical Institute, University of Oxford, Oxford OX2 6GG, UK}
\author{Peter D. Howell}
\affiliation{Oxford Centre for Industrial and Applied Mathematics, Mathematical Institute, University of Oxford, Oxford OX2 6GG, UK}
\author{Howard A. Stone}
\affiliation{Department of Mechanical and Aerospace Engineering, Princeton University, Princeton, NJ 08544, USA}

\begin{abstract}
We consider the unsteady thin-film dynamics of a long bubble of negligible viscosity that advances at a uniform speed in a cylindrical capillary tube. The bubble displaces a viscous, nonwetting fluid, creating a thin film between its interface and the tube walls. The film is considered thin enough that intermolecular forces in the form of van der Waals attractions are significant and thin-film rupture is possible. In the case of negligible intermolecular forces, a steady-state solution exits where a film of uniform thickness is deposited in the annular region between the bubble interface and the tube walls. However, once intermolecular interactions are important, the interface is perturbed out of its steady state and either (i) the perturbation grows sufficiently before reaching the rear meniscus of the bubble such that rupture occurs; or (ii) the perturbation remains small due to weak intermolecular forces until it leaves the bubble interface through the rear meniscus. We obtain, both numerically and asymptotically, the time-scale over which rupture occurs, and thus, we find a critical capillary number, depending on the bubble length and the strength of the intermolecular forces, below which the film is predicted to rupture.
\end{abstract}
\maketitle

\section{Introduction} \label{sec:intro}
When a bubble of negligible viscosity advances in a circular tube filled with a wetting, viscous fluid, a uniform thin film is deposited between the bubble interface and the tube walls. This setup was first studied experimentally 
\cite{fairbrother}, 
and 
it was shown that the speed of the bubble is higher than the average speed of the surrounding fluid \cite{Taylor}; this is an indication that the thickness of the deposited film increases with increasing bubble speed. A theoretical framework using the lubrication approximation was proposed by Bretherton \cite{bretherton}, who found that the thickness of the 
film varies as a two-thirds power law with the capillary number, which measures the relative strength of viscous to surface tension forces. 
This theoretical result was shown to be valid only in the limit of small capillary numbers, 
which was confirmed by experimental measurements in \cite{bretherton}.

Following Bretherton's contributions, there has been extensive theoretical and experimental work on the problem of a bubble of either negligible or finite viscosity moving in a tube and displacing a viscous, wetting fluid \citep{goldsmith, schwartzPrincen, Wong1, Wong2, parkhomsy, hodgesRallison, Lac2009, cantat}. Experimental results \cite{chen1985}, which were later verified by both theory and numerics \cite{chaudhury}, showed the range of capillary numbers for which Bretherton's theory is valid for an inviscid bubble, and below which the film thickness levels off and becomes independent of the capillary number.
%
Effects of nonwettability were first addressed by \cite{teletzke}, who showed numerically that there is a critical capillary number below which steady films of uniform thickness cannot be obtained due to rupture of the film. This result was qualitatively verified by experiments \cite{chen}. Other studies of the bubble in a tube problem have accounted for, e.g., flexibility of the substrate, which is an important effect in medical applications such as airway closure \citep{HalpernGaver, GaverHalpern, HazelHeil, Heil1999, Heil2000, Heil2001, Grothberh2010}. 

\begin{figure}[htb] 
\centering
\includegraphics[width=5in]{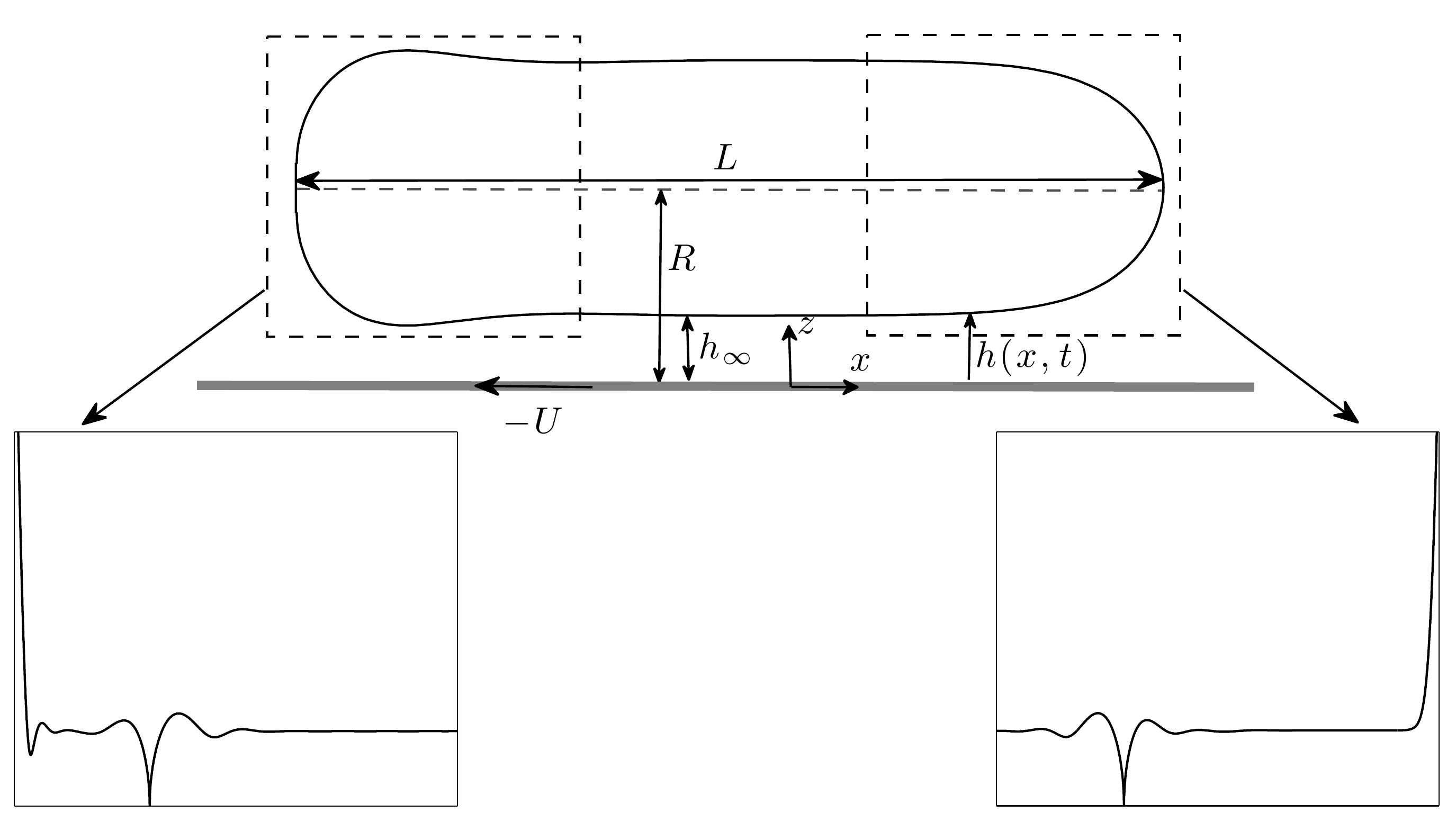}
\put(-190,200){\small(a)}
\put(-380,90){\small(b)}
\put(-130,90){\small(c)}
\caption{{\label{schematic} Bubble translating past a boundary. (a) Bubble interface with two domains highlighted: the front meniscus, and the rear meniscus of the bubble. (b) Rupture of the thin film as a perturbation advances away from the front meniscus. (c) Rupture of the thin film as a perturbation advances towards the rear meniscus.}}
\end{figure}

In this paper, we are interested in the particular situation of a nonwetting fluid surrounding the bubble, where disjoining pressure in the form of attractive intermolecular forces can be destabilizing (see, e.g., \citep{teletzke} for a review of the various disjoining pressure functions), and consequently,  the thin film may rupture. We are motivated by the experimental work of Chen \emph{et al.} \cite{chen}, who studied the dynamics of oil droplets advancing in a rectangular microchannel filled with water, where one region of the channel was made hydrophobic, while the other was kept hydrophilic. Rupture was observed in the hydrophobic region for slowly moving drops. However, once the drops were advanced at speeds larger than some critical value, it was observed that rupture was suppressed. 
To this end, we posit a simplified mathematical model to study the transition of a bubble of negligible viscosity from a wetted to a nonwetted region in a circular tube. We solve the model both analytically and numerically, and our results provide conditions under which rupture suppression is attainable.
The bubble configuration is illustrated in Fig.~\ref{schematic}(a); 
the subfigures \ref{schematic}(b) and \ref{schematic}(c) show two instances where rupture has occurred near the front or the rear meniscus of the bubble. The surface profiles are plotted from time-dependent numerical simulations of the front and rear menisci. Here, they are shown without scale only so as to illustrate the qualitative phenomena of rupture; see Sec.~\ref{s:unsteadySolns} for more details on the numerical implementation.

Note that rupture, or adhesion, is not always a desired phenomenon, especially in applications that involve self-cleaning surfaces \citep{shp3, shp1, shp2}. However, in some circumstances, rupture can be beneficial, such as in targeted drug delivery \citep{TDD}, or by using the adhesive properties of tumor cells to promote separation from healthy cells \citep{Blackstone}. The theoretical approach to thin-film rupture is typically developed in the framework of slow viscous flow (lubrication) theory, by the addition of attractive intermolecular forces, such as van der Waals attractions \citep{RuckensteinJain, BurelbachBankoff, IdaMiksis, WitelskiBernoff2000}. Self-similar analyses have been provided of rupturing film profiles \citep{WitelskiBernoff1999, zhang}. Also, studies have been reported of rupture delay by adding surfactants or increasing the flexibility of the substrate supporting the rupturing film \citep{MatarKumar}, and of rupture suppression by adding an external shear to the thin film \citep{Kalpathy, davis}.

We begin this study by developing the theoretical framework in Sec.~\ref{model} for a bubble advancing in a tube, and displacing a nonwetting fluid, where we use lubrication theory to describe the bubble dynamics when attractive van der Waals interactions are substantial. In Sec.~\ref{steady-state}, we study the steady-state profiles of the shape of the bubble as a function of a nondimensional van der Waals parameter that characterizes the strength of the intermolecular forces. In Sec.~\ref{asymptotics}, the asymptotic behavior of the steady states is analyzed for small and large values of the van der Waals coefficient, and a relation between film thickness and capillary number is found in Sec.~\ref{sss:mbr}. We analyze the unsteady dynamics in Sec.~\ref{s:unsteadySolns}; we begin with the case where van der Waals forces are significant throughout the tube and we perform a linear stability analysis as well as a numerical study in Sec.~\ref{ss:impExp}.
We then consider the problem of a bubble displacing a wetting fluid, when suddenly attractive van der Waals forces are turned on. The effect on the bubble dynamics is examined as the interface moves from a wetted to a nonwetted region in the tube, and is analyzed both numerically and asymptotically in Sec.~\ref{ss:disjoiningPress}. We conclude by finding a critical capillary number beyond which rupture is expected to be suppressed.


\section{Mathematical formulation} \label{model}
Consider a long bubble of negligible viscosity moving at a constant speed $U$ through a capillary tube of radius $R$, which is filled with a nonwetting fluid of viscosity $\mu$ and density $\rho$. The bubble is assumed to be at least a few tube radii long. The motion of the bubble displaces the viscous fluid, which causes the deposition of a uniform thin film in the annular region between the bubble interface and the tube walls [cf. Fig.~\ref{schematic}(a)]. The interfacial tension between the displaced fluid and the bubble is denoted by $\gamma$. The deposited film has a thickness $h_\infty$, and is considered thin enough that intermolecular forces in the form of long range van der Waals attractions are significant, and are measured by the Hamaker constant $A$.

In the limit that the film thickness is very small compared to the tube radius, the system can be described locally as two-dimensional. Thus, we use Cartesian coordinates $(x,z)$ and time $t$ to describe the spatial and temporal dynamics. The $x$-coordinate describes lateral positions of the bubble in the tube. We treat the front and rear menisci of the bubble separately, but shall discuss the implications of this assumption in Sec.~\ref{ss:impExp}. When treating the front meniscus of the bubble, we refer to $x\to\infty$ as the front bubble cap, $x\to -\infty$ as the uniform thin film of thickness $h_\infty$. Conversely, when treating the rear meniscus, we refer to $x\to -\infty$ as the rear bubble cap, $x\to \infty$ as the uniform thin film of thickness $h_\infty$.

Following Bretherton \cite{bretherton}, we assume that both the Weber number $\rho R U^2/\gamma$ and the Bond number $\rho g R^2/\gamma$ are very small, and we therefore neglect both inertial and gravitational effects. The thickness of the film from the tube wall, $z=0$, to the bubble interface is denoted by $h(x,t)$, and in the limit that $\abs{\partial_x h}\ll 1$, we can use the lubrication approximation to derive the equation describing thin-film dynamics (cf. Ref. \onlinecite{oron}) in the presence of attractive intermolecular forces. In dimensional form, the governing equation is 
\begin{equation}\label{dimEqdc}
{\partial_t h} + \frac{1}{3\mu}{\partial}_{x}
\left[\gamma h^3{\partial_{xxx}}{h} +
 \frac{A\mathcal{H}(x+Ut)}{2\pi h}{\partial_x}{h}\right] -  U\partial_x{h} = 0,
\end{equation}         
where $\mathcal{H}(x + Ut)$ will later be chosen (in Sec.~\ref{ss:disjoiningPress}) to be the Heaviside step function, so that a term proportional to $A$ switches on across $x = -Ut$ (in our reference frame moving with the advancing bubble). This will serve to model a section of the tube that suddenly transitions from wetting to nonwetting.

Throughout the remainder of this section and the next, we assume that $\mathcal{H} \equiv 1$ and hence the intermolecular forces apply throughout the entire length of the tube. 
Nondimensionalizing \eqref{dimEqdc} yields
\begin{equation}
\partial_T H + \partial_X \left[H^3 \partial_{XXX} H+ \frac{\beta}{H} \partial_X H \right] - \partial_X H = 0,
\label{dimensionlessFilmEq}
\end{equation}        
where we have set $h = h_\infty H$, $x = (\Ca^{-1/3}h_\infty)X$, $t = (h_\infty\Ca^{-1/3}U^{-1})T$, and where $\text{Ca}=3\mu U/\gamma$ is the capillary number. In \eqref{dimensionlessFilmEq}, we have also introduced the dimensionless van der Waals parameter,
\begin{equation}\label{beta}
\beta = \frac{A}{2\pi\gamma h_\infty^2 \text{Ca}^{2/3}}. 
\end{equation}
Note that when $\beta<0$, the intermolecular forces are repulsive, hence stabilizing, and the displaced fluid wets the substrate. When $\beta>0$, these forces are attractive, thus, the displaced fluid is nonwetting, and rupture may occur. In this work, we are interested in scenarios that may lead to rupture, and hence we consider the situation of a bubble surrounded by a nonwetting fluid in a tube, with $\beta>0$. 

The profile is initially set to 
\begin{equation}
H = H_s(X;\beta) \qquad \text{at $T = 0$}, \label{IC1}	
\end{equation}
where $H_s(X;\beta)$ describes the shape of either the front or rear meniscus at steady state, and will be discussed in more detail in Sec.~\ref{steady-state}. Our assumption is that, through the evolution, the film profile remains  quasi-static near the front or rear of the bubble, and hence we require that
\begin{subequations} \label{BCnew}
\begin{gather}
H \to 1 \quad \text{and} \quad \partial_X H \to 0 \quad  \text{as $X \to -\infty$ (front), $X \to +\infty$ (rear)} ,  \label{BCnewa} \\
H = H_s(X; \beta) + O(1) \quad \text{as $X \to +\infty$ (front), $X \to -\infty$ (rear)} . \label{BCnewb}
\end{gather}
\end{subequations}
%

In fact, we demonstrate in the next section that the relevant matching condition \eqref{BCnewb} requires that the curvature of the profile, $\partial_{XX} H$, matches $H_s'' \sim \kappa(\beta)$, corresponding to a linearized curvature of the meniscus of the bubble. 
Redimensionalizing, we see that $\kappa$ is related to the uniform film thickness $h_\infty$ in the bubble by
\begin{equation}\label{hinfOfBeta}
\frac{h_\infty}{R}=\kappa(\beta)\text{Ca}^{2/3}.
\end{equation}
The relation in \eqref{hinfOfBeta} provides an implicit description of how the film thickness varies as a function of the van der Waals coefficient $\beta$ for any $\beta>0$. In Sec.~\ref{asymptotics}, we show that in the limit of $\beta\to 0$, Bretherton's \cite{bretherton} principal result indicating the variation of the film thickness with $\Ca^{2/3}$ is retained as a special case of \eqref{hinfOfBeta}, with $\kappa = \kappa_0\approx0.643$.

\section{Steady-state theory ($\mathcal{H} \equiv 1$)} \label{steady-state}
Setting the time derivative in \eqref{dimensionlessFilmEq} to zero and integrating yields the governing steady-state problem for $H = H_s(X;\beta)$, 
\begin{equation}
H_s^3 H_s''' + \beta \frac{H_s'}{H_s} - H_s = -1,
\label{constFlux}
\end{equation}        
with primes ($'$) indicating differentiation in $X$.
The problem has been nondimensionalized so that the steady-state profile approaches a uniform meniscus of unit height in the central region; thus, $H_s \to 1$ as $X \to -\infty$ for the rescaled model at the front of the bubble, or $X \to \infty$ for the rear. Setting  $H_s = 1  + \eta(X)$, where $\abs{\eta} \ll 1$, we linearize \eqref{constFlux} to find the perturbation $\eta$ satisfies
\begin{equation} \label{etaequation}
\eta''' + \beta \eta' - \eta = 0.
\end{equation}
This equation has solutions of the form $\exp({m X})$, where $m$ is a root of the cubic equation $m^3 +\beta m -1 =0$. For $\beta>0$, this cubic equation has one positive real root, $m_1$, and two complex conjugate roots, $m_2$ and $m_2^*$, with negative real part, where * denotes complex conjugation. 

\begin{figure}[!t]
\centering
\includegraphics{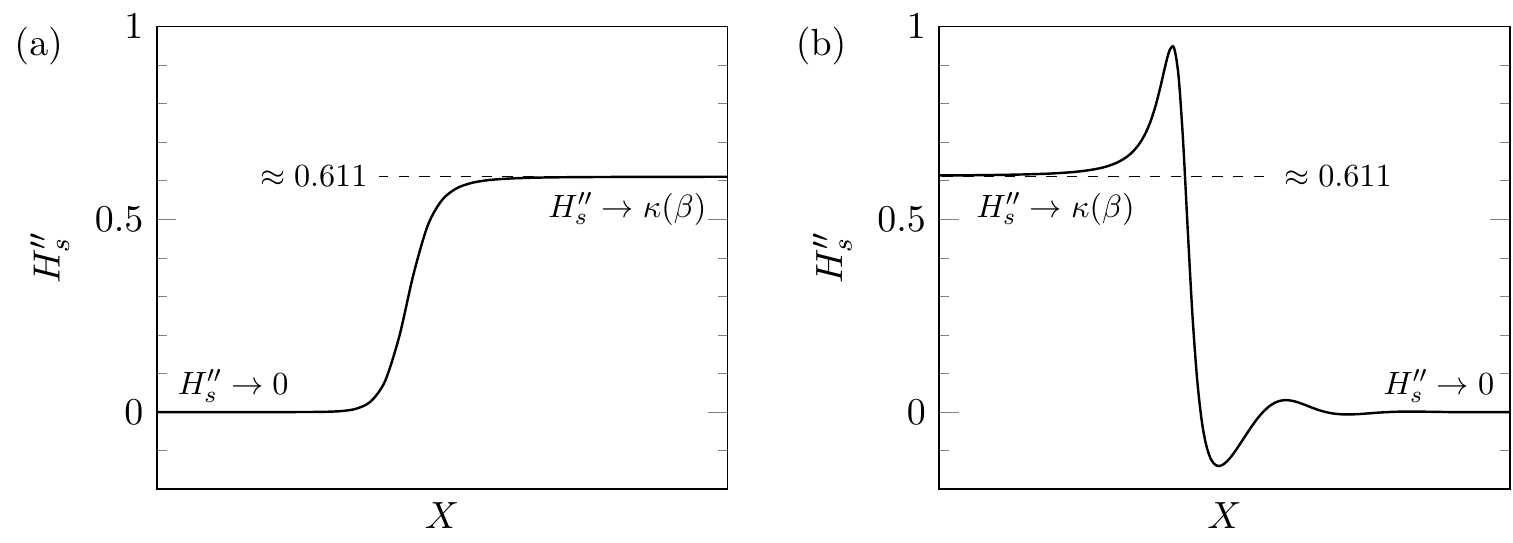}
\caption{Variation in linearized curvature of the bubble profile for the (a) front meniscus and (b) rear meniscus. The curvature tends to zero as the film tends towards the central region, and tends to a constant nonzero value $\kappa(\beta)$ towards the front (in a) and rear (in b). For the case of $\beta=0.2$, we find that $\kappa \approx0.611$. \label{bubbleProfile}}
\end{figure}

Solving for the perturbation $\eta$, we find the two limiting behaviors describing the matching of the front or rear of the bubble with the central body. At the front, in order to satisfy the uniform boundary condition, the modes corresponding to the roots $m_2$ and $m_2^*$ must equal zero. Thus, the solution asymptotes to a uniform film as
\begin{equation}
\text{(front meniscus)} \quad H_s(X; \beta) \sim 1+ \eta_0\exp({m_1 X}) \quad \text{ as $X\to -\infty$}, \label{eps0}	
\end{equation}
where $\eta _0$ is an integration constant, which may be set to unity without loss of generality by exploiting translation invariance. The numerical solution for the front meniscus of the bubble is found by solving \eqref{constFlux} subject to the farfield behavior \eqref{eps0}. A typical solution of the curvature $H_s''$ of the front meniscus is shown in Fig.~\ref{bubbleProfile}(a) for the case of $\beta=0.2$. Evidently, the curvature approaches a constant value as $X \to +\infty$, and we denote that value as $\kappa(\beta)=H_s''(+\infty)$. 

\begin{figure}[!t]
\centering
\begin{tikzpicture}
\node at (0,0) {\includegraphics[width=3.2in]{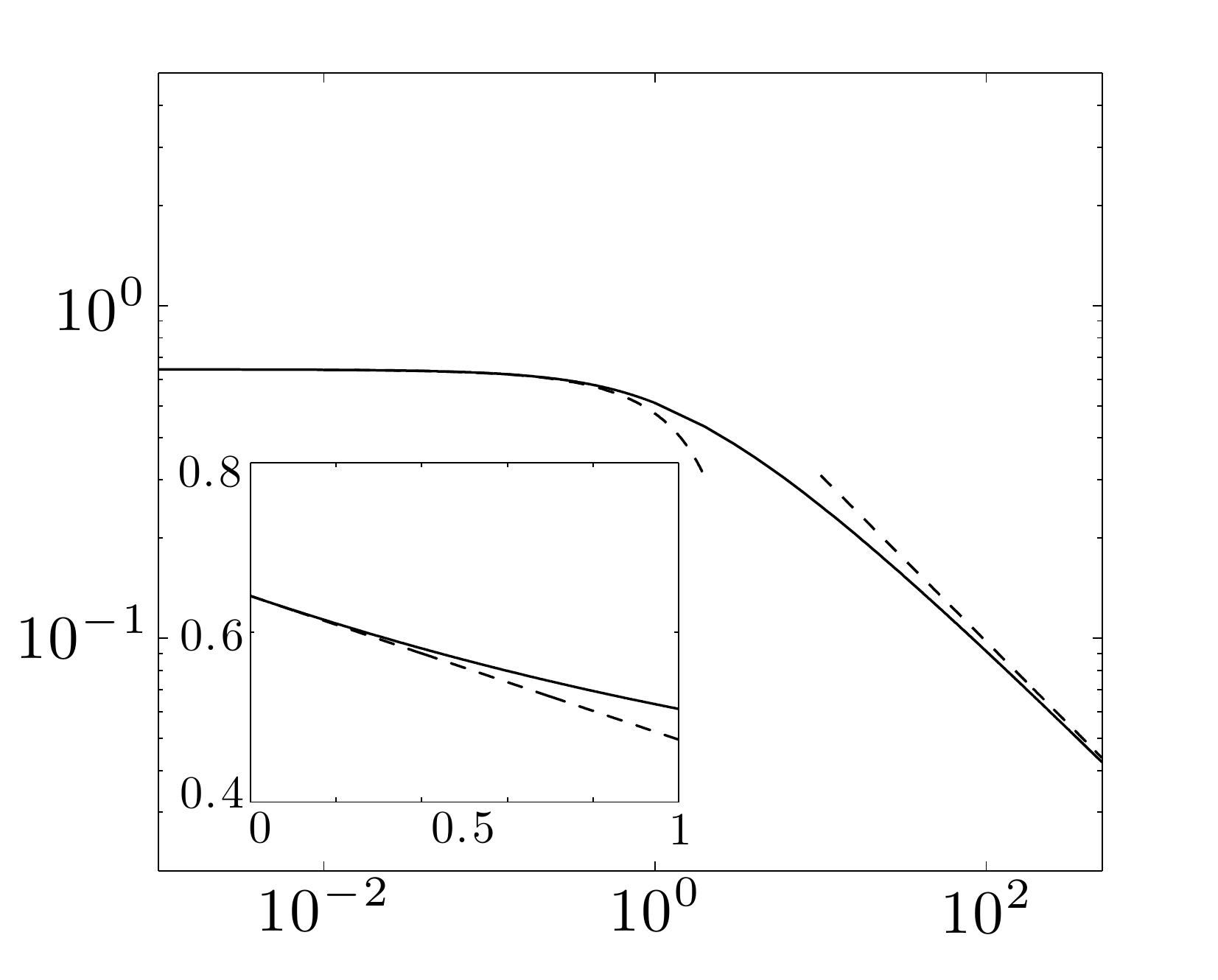}};
\node[scale=1] at (0.6,-3.2) {$\beta$};
\node[scale=1] at (-4.4,0.3) {$\kappa$};
\node[scale=1] at (-1.8,1.1) {$\beta\ll 1$};
\node[scale=1] at (2.7,-0.3) {$\beta\gg 1$};
\end{tikzpicture}
\caption{\label{kappaVSbeta} Numerically calculated curvature of the front meniscus of the bubble as a function of the dimensionless van der Waals interaction parameter $\beta$ (shown solid). The asymptotic predictions (shown dashed) in the small and large $\beta$ limits, given by \eqref{curvBetaSmall} and \eqref{curvBetaLarge}, closely match the numerical solutions. The inset shows the curvature as a function of $\beta$ on a linear scale for $\beta \leq 1$.}
\end{figure}
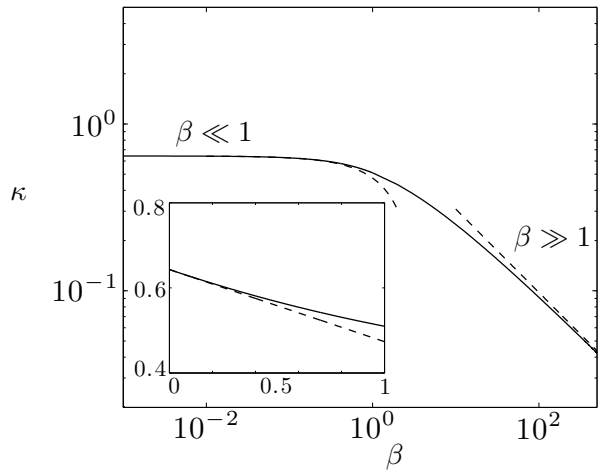

At the rear meniscus, in order to satisfy the boundary conditions, the eigenfunction of \eqref{etaequation} corresponding to $m_1$ must equal zero. Therefore, the rear meniscus asymptotes to a uniform film as
\begin{equation}
\text{(rear meniscus)} \quad H_s(X;\beta) \sim 1 + \tilde \eta_0\exp[{\text{Re}(m_2)X}] \cos[\text{Im}(m_2)X +\phi] \quad \text{as $X\to + \infty$},\label{pertubSolnRear}
\end{equation}
where the oscillations correspond to capillary ripples at the rear cap of the bubble \cite{bretherton, SDRWilson}. Here $\tilde \eta_0$ and $\phi$ are integration constants, and again $\tilde \eta_0$ may be set to 1 without loss of generality. Therefore, we use $\phi \in [-\pi, \pi]$ as a shooting parameter, and find the value $\phi^*$ of $\phi$ such that
the curvature of the rear meniscus of the bubble is equal to that of the front, $\kappa(\beta)$. In Fig.~\ref{bubbleProfile}(b) we show the curvature $H_s''$ of the rear meniscus for $\beta=0.2$. 
The behavior of the curvature as a function of $\beta$ is summarized in Fig.~\ref{kappaVSbeta}.

\subsection{Asymptotic Behavior of $\kappa(\beta)$} \label{asymptotics}
When $\beta=0$, Bretherton \cite{bretherton} showed that $h_\infty/R \sim \kappa_0 \text{Ca}^{2/3}$ [cf. Eq.~\eqref{hinfOfBeta}]. However, when $\beta>0$, the variation of the meniscus curvature $\kappa(\beta)$ leads to a modification of Bretherton's result. In what follows, we elucidate this effect by determining $\kappa$ in the limits of large and small $\beta$. Since the front meniscus sets the curvature $\kappa(\beta)$, which is then imposed on the rear meniscus, it is only necessary to perform our analysis on the front of the bubble.


In the limit of $\beta \to 0$, we expand $H_s = H_0 + \beta H_1 + O(\beta^2)$. Substituting the perturbation expansion into the governing equation \eqref{constFlux}, we obtain for the first two orders,
\begin{subequations}
\begin{gather}
H_0''' = (H_0-1)H_0^{-3}, \label{h0eqn} \\
H_1'''+(2 H_0^{-3} - 3H_0^{-4})H_1 = -H_0' H_0^{-4}, \label{h1eqn}
\end{gather}
\end{subequations}
%
subject to the boundary condition \eqref{eps0}, or $H_0\sim 1+\eta_0\exp(X)$ and $H_1 \to 0$ in the limit $X\rightarrow-\infty$. The leading-order equation in \eqref{h0eqn} is Bretherton's original equation (also known as the Landau-Levich equation). Numerical solutions for $H_0$ and $H_1$ confirm that the curvature at the front is given asymptotically by
\begin{align}\label{curvBetaSmall}
\kappa(\beta) = \lim_{X\to\infty} H_s''(X;\beta)\big|_{\rm front}
&\sim \kappa_0+\kappa_1\beta + \ldots \quad\text{as }\beta\rightarrow0,
\end{align}
where $\kappa_0 \approx 0.643$ and $\kappa_1 \approx -0.169$. 
Thus, when $\beta=0$ we retain Bretherton's result, with $\kappa=\kappa_0$.


We next turn our attention to the limiting behavior of $\kappa$ as $\beta\rightarrow\infty$. To reincorporate the spatial variation in this limit, we introduce the stretched coordinate $\tilde X = X/\beta$, with $H_s = H_\text{out}(\tilde{X})$. The governing equation \eqref{constFlux} then becomes
\begin{equation}
\frac{1}{\beta^3} H_\text{out}^3 H_\text{out}''' +  \frac{H_\text{out}'}{H_\text{out}}+ 1 - H_\text{out} = 0,
\label{largeBetaEq1}
\end{equation}
with primes now denoting derivatives with respect to $\tilde X$. Neglecting the first term of \eqref{largeBetaEq1} yields the leading-order result,
\begin{equation} \label{Houter}
H_{\text{out}} \sim
\left[ 1-\exp(\tilde X- \tilde X_0)\right]^{-1},
\end{equation}
where the integration constant $\tilde X_0$ corresponds to an arbitrary translation in $\tilde X$.
We shall refer to \eqref{Houter} as the outer solution, which automatically satisfies the farfield condition $H_\text{out} \to 1$ as $\tilde{X} \to -\infty$. However, the asymptotic solution \eqref{Houter} predicts blowup of the film thickness, with $H_{\text{out}}\rightarrow\infty$ as $\tilde{X}\rightarrow\tilde{X_0}$. We therefore seek a boundary layer near the transition point $\tilde{X} = \tilde{X_0}$, in which the dominant balance in \eqref{largeBetaEq1} changes and incorporates the surface tension term.

We refer to the solution near $\tilde{X}=\tilde{X_0}$, as the inner solution. Here, we perform the rescalings $\tilde{X} = \tilde{X_0} + \beta^{-1/2}\hat X$ and $H_{\text{out}}(\tilde X) = \beta^{1/2} H_\text{in}(\hat X)$,
which transforms \eqref{largeBetaEq1} into
\begin{equation}
H_\text{in}^3 H_\text{in}''' + \frac{H_\text{in}'}{H_\text{in}}- H_\text{in} =-\frac{1}{\sqrt{\beta}},
\label{largeBetaEq2}
\end{equation}
with primes now denoting differentiation with respect to $\hat X$. Then, as $\beta \to \infty$, we express the inner solution as an asymptotic expansion of the form $H_{\text{in}} \sim H_{\text{in}0}+\beta^{-1/2}H_{\text{in}1}+O(\beta^{-1})$. The leading- and first-order solutions satisfy the differential equations
\begin{subequations}\label{HinODEs}
\begin{gather}
H_{\text{in}0}^3 H_{\text{in}0}'''+\frac{H_{\text{in}0}'}{H_{\text{in}0}}
-H_{\text{in}0}=0,
\\
H_{\text{in}0}^3H_{\text{in}1}'''+\frac{H_{\text{in}1}'}{H_{\text{in}0}}
+\left(2-\frac{4H_{\text{in}0}'}{H_{\text{in}0}^2}\right)H_{\text{in}1}
=-1,
\end{gather}
\end{subequations}
while matching with the outer solution \eqref{Houter} leads to the farfield conditions
\begin{equation} \label{Hin_match}
H_\text{in0} (\hat X) 
\sim-\frac{1}{\hat X}+\frac{6}{5\hat{X}^7}+\cdots,
\quad
H_\text{in1} (\hat X) 
\sim\frac{1}{2}-\frac{3}{\hat{X}^6}+\cdots,
\quad\text{as }\hat{X} \rightarrow-\infty.
\end{equation}
The inner solutions $H_\text{in0}$ and $H_\text{in1}$ are numerically computed by solving \eqref{HinODEs} as an initial-value problem subject to the farfield conditions \eqref{Hin_match} applied at large negative $\hat{X}$. Then, the differential equations \eqref{HinODEs} are integrated to a large positive value of $\hat X$ such that the curvatures $H_\text{in0}''$ and $H_\text{in1}''$ reach constant values $\kappa_\text{in0}$ and $\kappa_\text{in1}$, up to numerical error. Hence, the large-$\beta$ asymptotic behavior of the function $\kappa(\beta)$ is found to be given by
\begin{equation} \label{curvBetaLarge}
\kappa
\sim \frac{\kappa_\text{in0}}{\beta^{1/2}}+\frac{\kappa_\text{in1}}{\beta}+\ldots \qquad \text{ as $\beta\rightarrow\infty$},	
\end{equation}
where $\kappa_\text{in0}\approx~0.977$ and $\kappa_\text{in1}\approx-0.657$.
The asymptotic results \eqref{curvBetaSmall} and \eqref{curvBetaLarge} are displayed as dashed lines in Fig.~\ref{kappaVSbeta}, and are in excellent agreement with the numerical solution of the original differential equation \eqref{constFlux}.

\subsection{A modification of the Bretherton relation}\label{sss:mbr}

In Bretherton's work \cite{bretherton}, corresponding to $\beta=0$, the normalized film thickness $h_\infty/R$ scales with $\text{Ca}^{2/3}$.
Our generalized relation \eqref{hinfOfBeta} depends implicitly on both $h_\infty/R$ and $\text{Ca}$ through the van der Waals parameter $\beta$, defined by \eqref{beta}. To unravel the dependence of the deposited film thickness on the capillary number, it is helpful to define an additional dimensionless parameter
\begin{equation}\label{deltadef}
\ell=\left(\frac{A}{2\pi\gamma R^2}\right)^{1/2},
\end{equation}
which corresponds to a molecular length scale that depends only on the physical properties of the fluid and the tube radius $R$.
Typical values of the Hamaker constant $A$ are in the range $10^{-20}$--$10^{-19}$\,J, the interfacial tension $\gamma$ has an order of magnitude of about $10^{-2}\text{J}/\text{m}^2$, and $R$ may be from a few to a few hundred microns. Therefore, $\ell$ is typically small, taking values in the range $10^{-6}$ to $10^{-3}$.

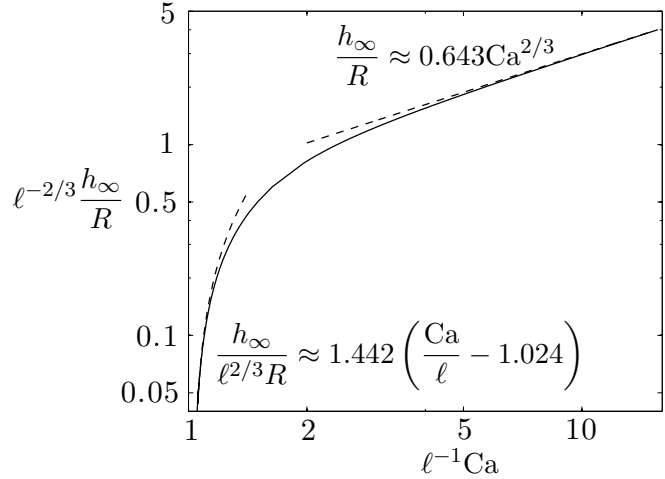
\begin{figure}
\centering
\begin{tikzpicture}
\node at (0,0) {\includegraphics[width=3.2in]{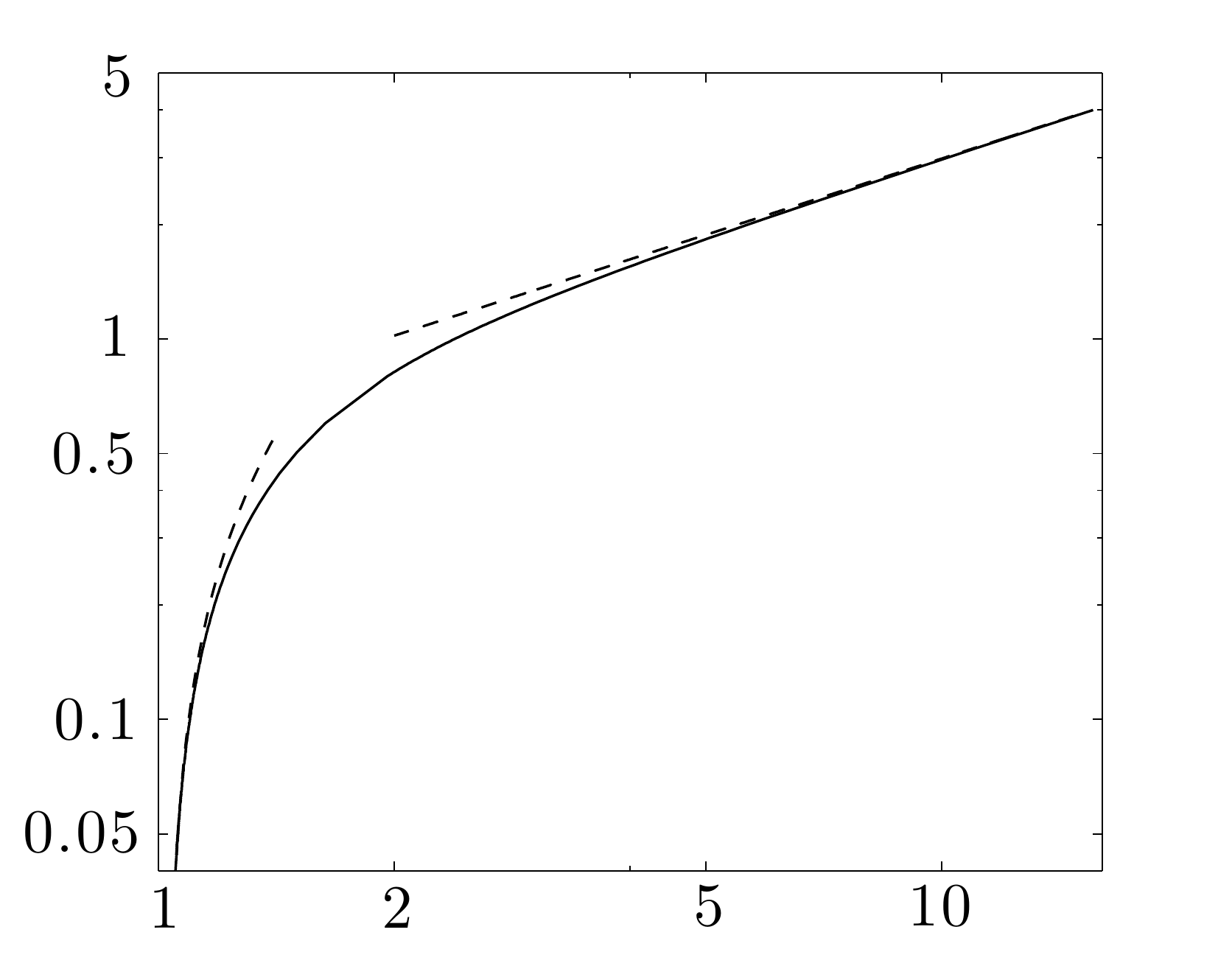}};
\node[scale=1] at (0.6,-3.2) {$\ell^{-1}\text{Ca}$};
\node[scale=1] at (-4.6,0.3) {$\displaystyle \ell^{-2/3}\frac{h_\infty}{R}$};
\node[scale=1] at (-0.2,-1.8) {$\displaystyle \frac{h_\infty}{\ell^{2/3}R} \approx 1.442 \left( \frac{\Ca}{\ell} -1.024\right)$};
\node[scale=1] at (0.4,2.2) {$\displaystyle \frac{h_\infty}{R} \approx 0.643 \Ca^{2/3}$};
\end{tikzpicture}
\caption{The relation between the normalized film thickness $\ell^{-2/3}h_\infty/R$ and the scaled capillary number $\ell^{-1}\text{Ca}$, given parametrically by
\eqref{hinfCapara}. The dashed curves show the asymptotic limits
\eqref{Cahinfbto0} as $\beta\rightarrow0$ and \eqref{Cahinfbtoinfty} as $\beta\rightarrow\infty$.}
\label{hinfvsCa}
\end{figure}

We use \eqref{beta} and \eqref{hinfOfBeta} to express both $h_\infty/R$ and $\text{Ca}$ as functions of $\ell$ and $\beta$, namely,
\begin{equation}
\label{hinfCapara}
\ell^{-1}\text{Ca} =\frac{1}{\beta^{1/2}\kappa(\beta)}
\quad \text{and} \quad
\ell^{-2/3}\left(\frac{h_\infty}{R}\right)=
\left(\frac{\kappa(\beta)}{\beta}\right)^{1/3}.	
\end{equation}
As $\beta$ ranges between $0$ and $\infty$, equations \eqref{hinfCapara} parametrically define a functional relationship between $\ell^{-2/3}h_\infty/R$ and $\ell^{-1}\text{Ca}$, which is plotted as a solid curve in Fig.~\ref{hinfvsCa}.
This indicates that the film thickness $h_\infty$ is an increasing function of the capillary number, as expected.
The detailed behavior may be clarified by using the asymptotic approximations of $\kappa(\beta)$ obtained above. We point out that the bubble must always fit in the tube, and therefore the increase of the film thickness with capillary number must be modified for large values of $\Ca$ \citep{AussillousQuere, Klaseboer2014}. This, however, is not analyzed in the current study.

Considering the small and large limits of $\beta$, and combining \eqref{curvBetaSmall} and \eqref{curvBetaLarge} with \eqref{hinfCapara}, we obtain
\begin{subequations}
\begin{align}
\ell^{-1}{\Ca} &\sim\frac{1}{\kappa_0}\beta^{-1/2},
&
\ell^{-2/3}\frac{h_\infty}{R}&\sim{\kappa_0^{1/3}}\beta^{-1/3}
\sim\kappa_0\left(\frac{\text{Ca}}{\ell}\right)^{2/3}
&&\text{for $\beta \to 0$}, \label{Cahinfbto0} \\
\ell^{-1}{\Ca}&\sim\frac{1}{\kappa_\text{in0}}
+\frac{\left|\kappa_\text{in1}\right|}{\kappa_\text{in0}^2}\,
\beta^{-1/2},
&
\ell^{-2/3}\frac{h_\infty}{R}&\sim\kappa_\text{in0}^{1/3}\beta^{-1/2}
&&\text{for $\beta \to \infty$}, \label{Cahinfbtoinfty}
\end{align}
\end{subequations}
where $\kappa_0\approx0.643$, $\kappa_{\rm in 0} \approx 0.977$, and $\kappa_\text{in1}\approx-0.657$. The latter equation \eqref{Cahinfbtoinfty} can be rearranged to 
\begin{equation} \label{Cahinfbtoinfty2}
\frac{h_\infty}{\ell^{2/3}R} \sim
\frac{\kappa_\text{in0}^{7/3}}{\left|\kappa_\text{in1}\right|}
\left(\frac{\text{Ca}}{\ell}-\frac{1}{\kappa_\text{in0}}\right) 
\approx
1.442
\left(\frac{\text{Ca}}{\ell}-1.024\right).
\end{equation}
The dashed curves in Fig.~\ref{hinfvsCa} verify that the approximate relations \eqref{Cahinfbto0} and \eqref{Cahinfbtoinfty2} are approached in the relevant limits.

Now, we can interpret the results as follows: First, Bretherton's result is recovered when $\beta$ is small, which corresponds to the capillary number being sufficiently large, namely, 
\begin{equation}
\ell\ll\text{Ca}\ll1. 
\end{equation}
Second, the relation between $h_\infty/R$ and $\text{Ca}$ starts to depart significantly from Bretherton's prediction when $\text{Ca}=O(\ell)$. Finally, the film thickness $h_\infty$ left behind the front meniscus reaches zero at a finite critical value of the capillary number,
\begin{equation}
\text{Ca}_\text{crit}\approx1.024\,\ell.
\end{equation}
Thus, for capillary numbers smaller than this value, no steady film will be left behind the advancing front meniscus. This result differs quite significantly from the case where the surrounding fluid wets the substrate ($\beta\leq 0$). In that case, the departure from Bretherton's results for $\Ca = O(\ell)$ is characterized by a uniform film independent of capillary number \citep{chen1985, chaudhury}.


\section{Unsteady Solutions} \label{s:unsteadySolns}

\noindent So far, we have shown that for a bubble moving in a tube and subject to attractive van der Waals forces, steady-state free-surface profiles exist for $\beta > 0$. The next question is whether such steady-state solutions remain stable or become unstable 
once they are evolved in time.
Our efforts in this section are divided into two parts. In Sec.~\ref{ss:impExp} the stability of the bubble is studied using numerical simulations of the central thin-film and meniscus regions, where van der Waals forces are assumed to be substantial throughout the entire length of the tube. Here we shall confirm the dependence of the rupture time $T_R$ on the van der Waals parameter $\beta$. Then, in Sec.~\ref{ss:disjoiningPress} we shall propose a 
model 
where the bubble moves abruptly from a wetted region ($\beta = 0$) to a nonwetted region ($\beta > 0$) in the tube. Here we show that the rupture time differs from that obtained in the case where van der Waals forces are significant everywhere in the tube, and we find the conditions under which bubble rupture may be suppressed.

\subsection{Evolution of a bubble subject to van der Waals attractions everywhere $(\mathcal{H} \equiv 1$)} \label{ss:impExp}

Consider a situation where a bubble displaces a nonwetting fluid that is present throughout a tube. The evolution of the bubble is described by equation \eqref{dimensionlessFilmEq}, or equivalently equation \eqref{dimEqdc} with $\mathcal{H}\equiv 1$. The steady-state profiles $H_s(X,\beta)$ of this setup are analyzed in Sec.~\ref{steady-state}, and in this section we determine the stability of such profiles.

First, we study the linear stability of the time-dependent equation \eqref{dimensionlessFilmEq} using a normal mode analysis. Perturbing the base state of uniform film thickness, we set $H(X,T;\beta) = 1 + \epsilon \exp({\text{i}kX + \omega T})$ and linearize for infinitesimal perturbations in the limit $\epsilon\ll 1$. This yields the dispersion relation
\begin{equation} \label{dispersion}
\omega = -k^4 + \beta k^2 +\text{i}k,
\end{equation}
between the linear growth rate $\omega$ and the wave number $k$. The result is essentially the same as the dispersion relation obtained for a stationary film subject to viscous, surface tension, and van der Waals forces~\cite{davis} except for the addition of a traveling wave term $\text{i}k$. Thus, according to \eqref{dispersion}, linear theory predicts that the film is stable for $\beta\leq 0$ and unstable to long-wave perturbations for any $\beta>0$. Moreover, from \eqref{dispersion}, we find that the most rapidly growing mode, i.e., the wavenumber $k_c$ that maximizes $\text{Re}(\omega)$,
corresponds to a wavelength and growth rate given by
$\lambda_c ={2\pi}/{k_c} =  2\pi\sqrt{{2}/{\beta}}$, and 
$\Re(\omega_c)=-k_c^4+\beta k_c^2={\beta^2}/{4}$,	
respectively.
We conclude that wavelengths of order $\beta^{-1/2}$ or longer are expected to be unstable, and that the characteristic time for the instability to occur should be proportional to $\beta^{-2}$.

Next, we seek to determine the time required for the thin film between the bubble interface and tube wall to rupture, and furthermore, demonstrate how this process can be controlled by adjustments to $\beta$. We shall study the time evolution of a configuration that begins near the idealized steady-state profile $H_s(X;\beta)$, and as explained in Sec.~\ref{model}, our model assumes that the front and rear menisci are static during the instability regime and do not interact with one another. Thus, the evolution primarily occurs in the central region of the bubble. We return to discuss the possibility of numerical models of the full (two-dimensional) bubble in our discussion of Sec.~\ref{sec:discuss}.

We 
solve the time-dependent equation \eqref{dimensionlessFilmEq} subject to initial condition \eqref{IC1} and boundary conditions \eqref{BCnew}. 
An implicit-explicit numerical scheme is used, which computes nonlinear terms of $H$ in \eqref{dimensionlessFilmEq} explicitly, while the linear terms of $H$ and its derivatives are treated implicitly for increased stability.  Centered differences are employed in both the spatial and temporal coordinates. We use a uniform spatial grid with step size $\Delta X$ (typically between $10^{-2}$ and $10^{-3}$), and an adaptive time step that decreases with the growth of the maximum traveling-wave amplitude (typically between $10^{-3}$ and $10^{-8}$). Further details of this numerical scheme are presented in Appendix~B in \cite{thesis_NH} or Appendix~A in \cite{ren_2015_on_the}.

\begin{figure} [t]
\centering
\includegraphics[width=3in]{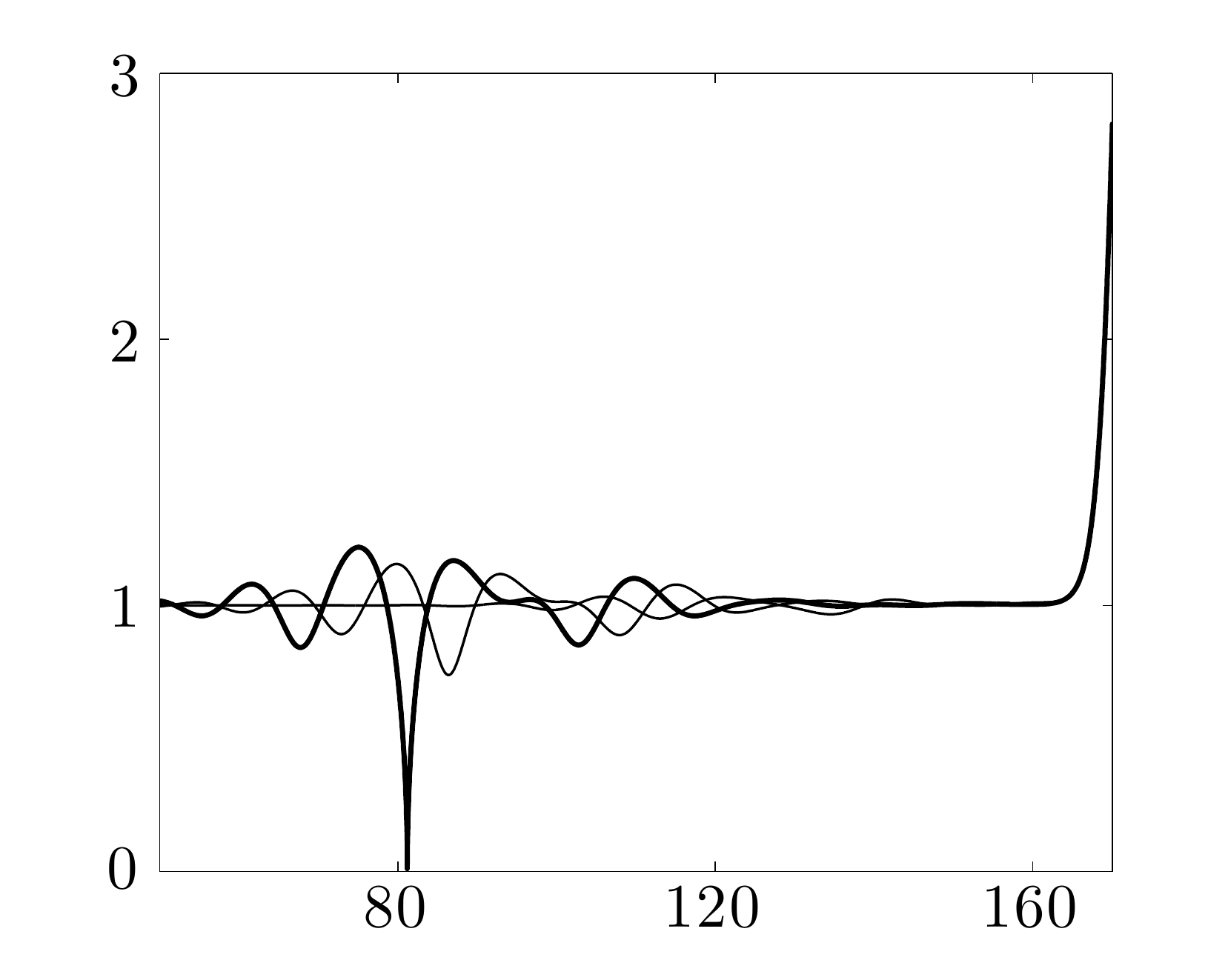}
\put(-220,70){\rotatebox{90}{{$H(X,T)$}}}
\put(-112,0){{$X$}}
\put(-220,155){(a)}
\quad\quad \quad\includegraphics[width=3in]{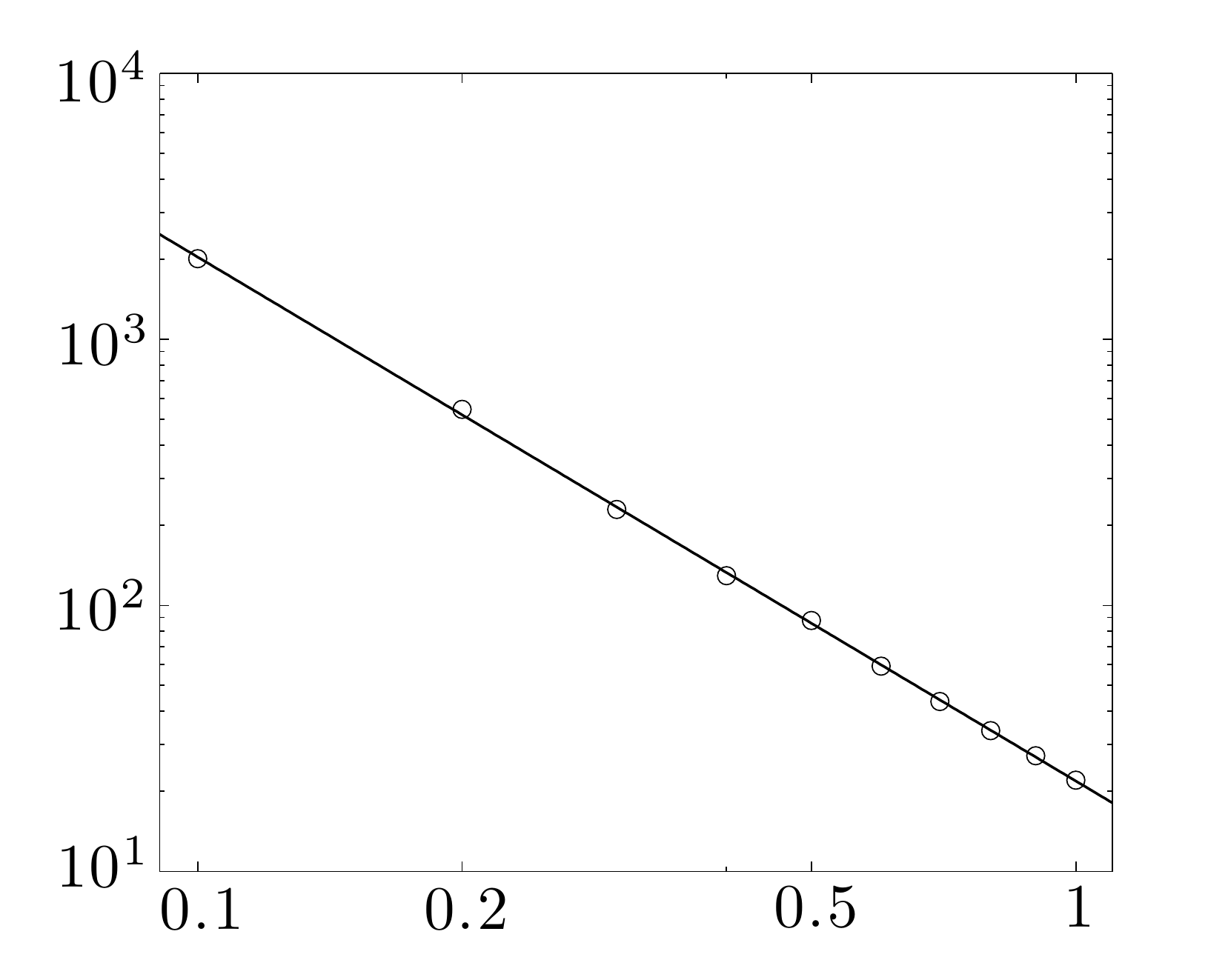}
\put(-230,95){{$T_R$}}
\put(-105,105){\small{$T_R\approx \beta^{-2}$}}
\put(-112,0){{$\beta$}}
\put(-230,155){(b)}
\caption{\label{ruptureTimeAndPositionVSbeta} (a) Evolution to rupture for the case of the central film near the front meniscus, with $\beta=0.5$ at $T=56.95, \,\, 83.65, \,\, 88.84$. Such simulations were used to create a graph of (b) the rupture time $T_R$ plotted against $\beta$. Within the range of $\beta$ shown, the rupture time is proportional to $\beta^{-2}$.}
\end{figure}
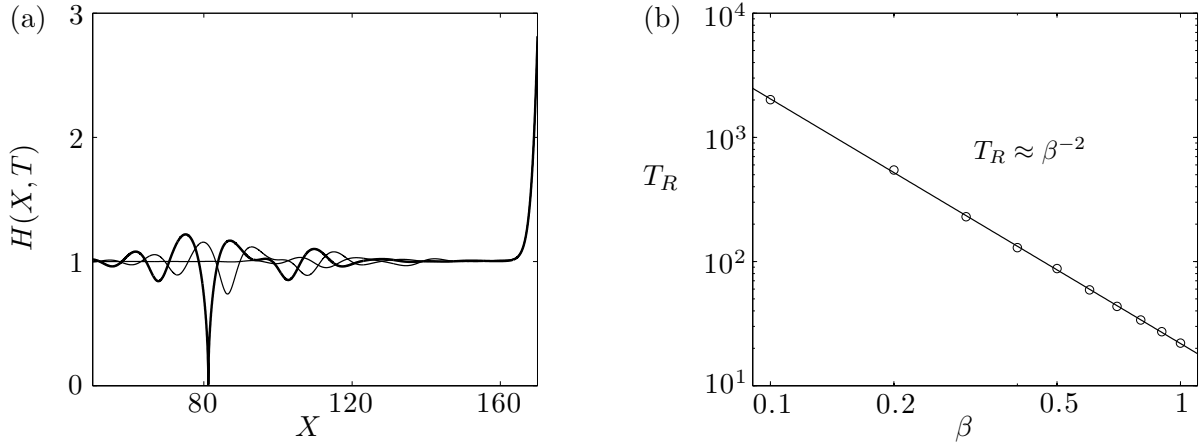

When the initial condition \eqref{IC1}, i.e., the steady-state  $H_s(X;\beta)$, is evolved in time using \eqref{dimensionlessFilmEq}, a perturbation starts near the front meniscus and grows in amplitude as it travels towards the central body of the bubble. If the amplitude grows sufficiently, rupture occurs. Fig.~\ref{ruptureTimeAndPositionVSbeta}(a) shows a typical result corresponding to evolution near the front meniscus for the case $\beta = 0.5$. Simulations are run for different values of $\beta \in [0.1, 1]$; for each run, we report the time it takes for rupture to occur, $T_R$, and this is shown in Fig.~\ref{ruptureTimeAndPositionVSbeta}(b). Hence, we find that $T_R$ varies proportional to $\beta^{-2}$ as predicted by linear stability theory. Thus, in regards to the model, as the van der Waals forces become negligible, i.e., $\beta\to 0$, rupture is suppressed with $T_R \to \infty$.

On the other hand, when the steady-state profile of the rear meniscus is evolved in time over the range of $\beta \in [0.1,0.5]$, it remains stable for the entire simulation time, which was at least an order of magnitude higher than the rupture time $T_R$ obtained from the front meniscus simulation for a given $\beta$. Additionally, even when a perturbation is manually introduced to this profile, we find that it is swept up by the rear bubble cap if its amplitude is small enough, i.e., if the amplitude does not grow sufficiently to lead to thin-film rupture before reaching the rear of the bubble.
This indicates that rupture can be suppressed as long as the wave perturbation reaches the rear bubble cap before the instability occurs. As we shall note in Sec.~\ref{sec:discuss}, a full simulation of this process would necessarily involve the finite-length nature of the bubble and modeling the interactive nature of the bubble's various regions.


Based on these numerical results, we summarize as follows: given a value $0<\beta\ll 1$, a traveling wave perturbation moves from the front meniscus towards the rear and grows until either (i) rupture occurs in the central region of the bubble; or (ii) the wave reaches the rear meniscus, where the van der Waals forces are minimal, and rupture is suppressed insofar as this process is concerned.

%


\subsection{Evolution of a bubble subject to van der Waals forces in one section of the tube} \label{ss:disjoiningPress}



\begin{figure}[t!] 
\centering
\includegraphics[width=4.5in]{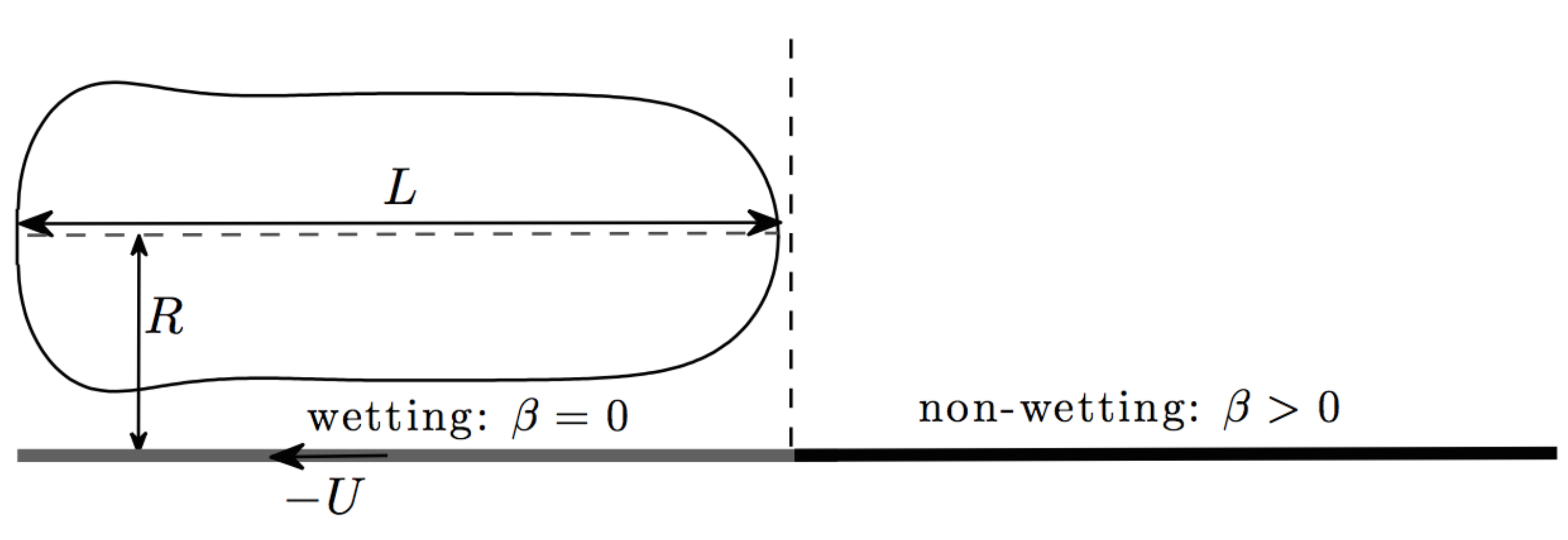}
\caption{\label{schem2} Schematic showing the section of a tube where the disjoining pressure is switched on. The bubble translates across a section of the surface where the wetting state changes (from wetting to nonwetting). This is modeled as a step transition in the thin-film equation \eqref{dimEqdc}.}
\end{figure}

As noted in Sec.~\ref{ss:impExp} above, when van der Waals forces are significant throughout the entire length of a tube, the time it takes for a bubble to rupture is proportional to $\beta^{-2}$. Now, we consider a situation where the destabilizing disjoining pressure is only present in one section of the tube (Fig.~\ref{schem2}), and we find how the rupture time varies as a function of the van der Waals coefficient $\beta$, which then motivates us to find the critical capillary number above which rupture is expected to be suppressed. We thus use the dimensional governing equation \eqref{dimEqdc} and choose $\mathcal{H}$ to be the Heaviside step function so that the term proportional to $A$ switches on across $x=-Ut$. Note that when $t$ is negative and before the front meniscus has reached $x=-Ut$, there is no disjoining pressure effect and the problem reduces to Bretherton's model \cite{bretherton}. In this case, the film thickness is given by \eqref{hinfOfBeta} with $\kappa = \kappa_0 \approx 0.643$, and we now use this thickness specifically when nondimensionalizing
\eqref{dimEqdc}. The definition \eqref{beta} of $\beta$ is replaced by a new dimensionless parameter
\begin{equation}
\tilde\beta=\frac{A}{2\pi\gamma R^2\kappa_0^2\Ca^2}
=\frac{\ell^2}{\kappa_0^2\Ca^2},
\end{equation}
where $\ell$ is still defined by \eqref{deltadef}. We also use the results of the linear stability analysis to select an appropriate time-scale over which the disturbances to the uniform thin film are expected to grow.
These ideas motivate the following choice of new dimensionless variables:
\begin{align}\label{smalland}
h &= ({\kappa_0R\,\Ca^{2/3}})\, \eta,
&
x &=({\kappa_0R\,\Ca^{1/3}})\, \xi,
&
t&=({\tilde\beta^{-2}\kappa_0^5R\,\Ca^{1/3}U^{-1}}) \tau.
\end{align}
Then, from \eqref{dimEqdc}, the governing system for the front meniscus becomes
\begin{subequations}
\begin{gather}\label{ndEqdc}
\tilde\beta^2\partial_\tau \eta +
\partial_\xi
\left[\eta^3 \partial_{\xi\xi\xi}\eta +
\tilde\beta\frac{\mathcal{H}(\tau+\tilde\beta^2\xi)}{\eta}
\partial_\xi\eta - \eta\right] = 0, \\
\eta\rightarrow\eta_\infty 
\text{ as $\xi\rightarrow-\infty$}, \qquad
\partial_{\xi\xi}\eta \rightarrow\kappa_0
\text{ as $\xi\rightarrow+\infty$},
\end{gather}
\end{subequations}
where $\eta_\infty$ is the normalized thickness of the deposited liquid film.

The parameter $\tilde\beta$ is small, and when terms of order $\tilde\beta^2$ are neglected, \eqref{ndEqdc} becomes quasi-steady. For $\tau<0$, we retrieve Bretherton's problem, for which we know that a uniform film of unit dimensionless thickness will be deposited, i.e., $\eta_\infty=1$. For $\tau>0$, the disjoining pressure term switches on, and for $\tilde\beta\ll1$  the resulting deposited film thickness may be inferred from the results of Sec.~\ref{sss:mbr}, where $\eta_\infty= \kappa(\tilde\beta)/\kappa_0 \sim 1- (|\kappa_1|/\kappa_0)\tilde\beta$ when the smallness of $\tilde\beta$ is exploited.
Thus, the thickness of the film deposited behind the front meniscus decreases by a small factor of order $\tilde\beta$ as $\tau$ increases through zero.


\subsubsection{Numerical solution in the traveling frame} \label{ss:numSoln}

To follow the progress of the instability, we shift to a frame that moves with the tube wall by using the traveling-wave coordinate
\begin{equation}
\zeta=\tilde\beta^{1/2}\left(\xi+\tilde\beta^{-2}\tau\right),
\end{equation}
which transforms \eqref{ndEqdc} to
\begin{subequations} \label{eq30}
\begin{gather} \label{ndEqtw}
\partial_\tau\eta +
\partial_\zeta
\left[\eta^3 \partial_{\zeta\zeta\zeta}\eta +
\frac{\mathcal{H}(\zeta)}{\eta}
\partial_\zeta\eta \right] = 0,
\end{gather}
\text{while matching with the quasi-static front meniscus implies the boundary and initial conditions,}
\begin{gather}\label{ndtwbcs}
\eta\rightarrow 1 \text{ as $\zeta\rightarrow-\infty$}, \qquad 
\eta\rightarrow 1-\nu \text{ as $\zeta\rightarrow+\infty$}, \qquad
\eta = 1-\nu\mathcal{H}(\zeta) \text{ at $\tau=0$},
\end{gather}
\end{subequations}
where we have introduced the shorthand
\begin{equation}
\nu=\frac{\left|\kappa_1\right|}{\kappa_0}\,\tilde\beta
=\frac{\left|\kappa_1\right|}{\kappa_0^3}\,\frac{\ell^2}{\Ca^2}
\approx0.636\,\frac{\ell^2}{\Ca^2}.
\end{equation}

\begin{figure} [t]
\centering
\begin{minipage}[t]{1.0\linewidth}
\includegraphics[width=3in]{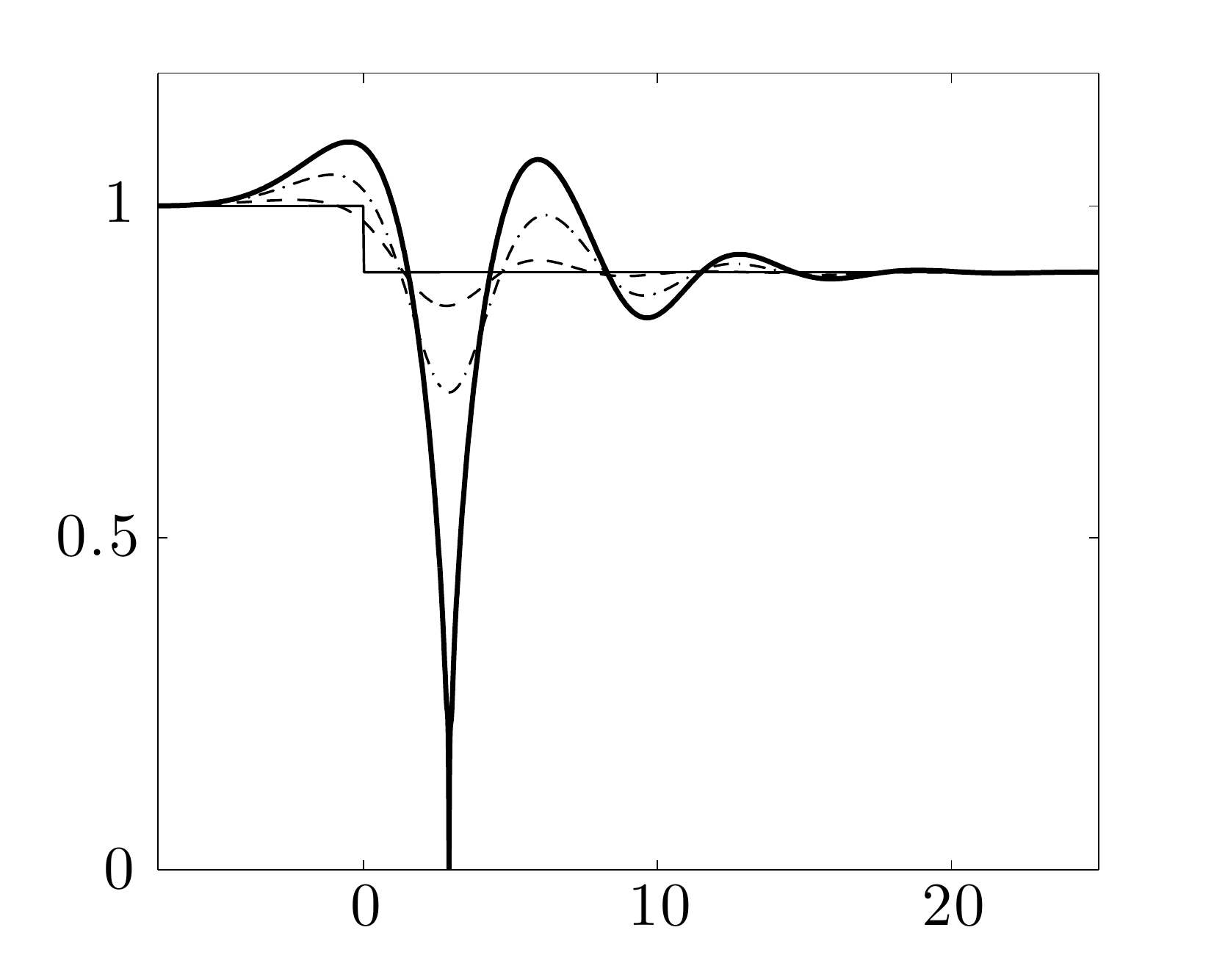}
\put(-225,95){{$\eta$}}
\put(-112,0){{$\zeta$}}
\put(-225,155){(a)}
\quad\quad\quad\includegraphics[width=3in]{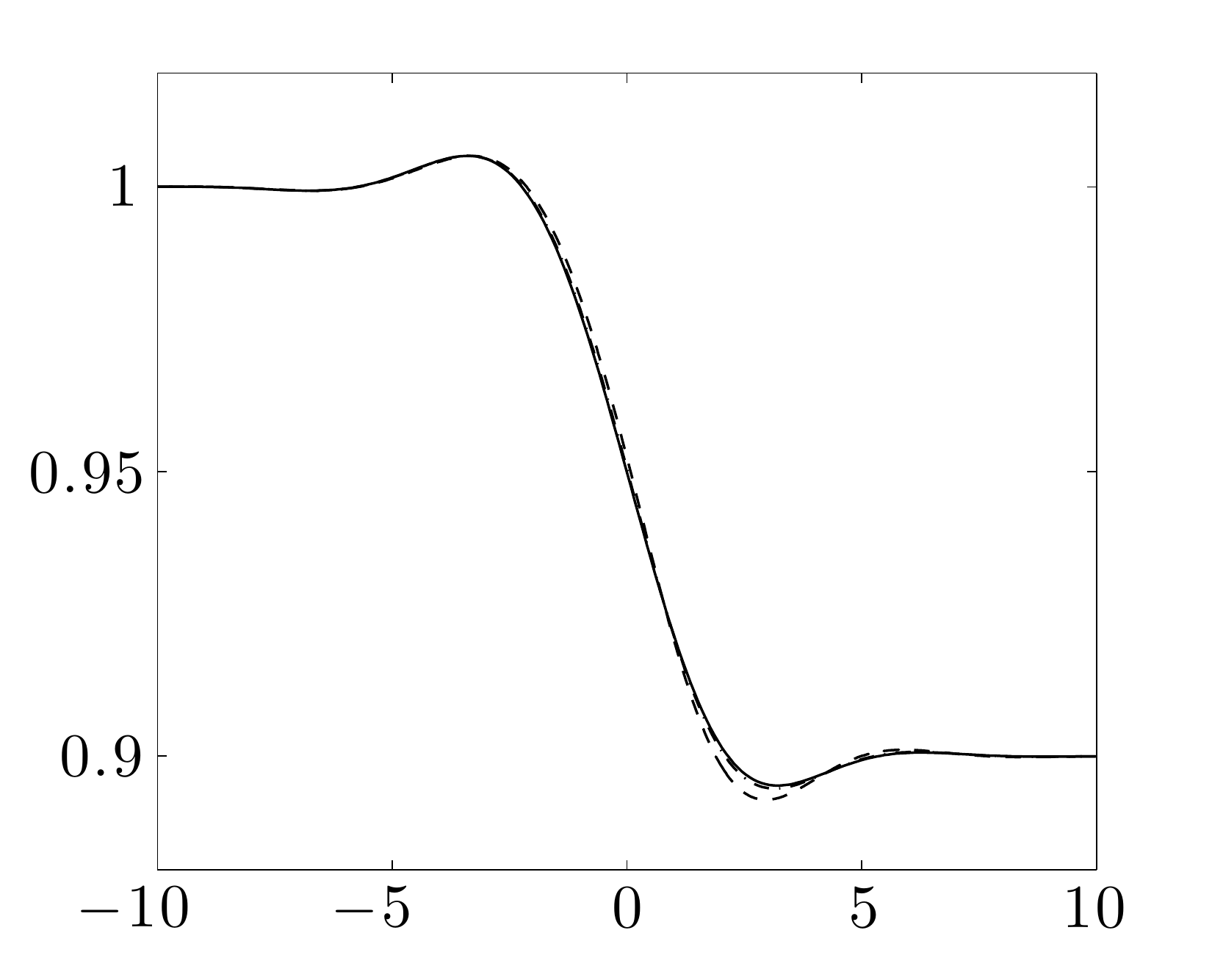}
\put(-225,95){{$\eta$}}
\put(-112,0){{$\zeta / \tau^{1/4}$}}
\put(-225,155){(b)}
\end{minipage}
\caption{\label{fig:mols1} (a) The numerical solution $\eta(\zeta,\tau)$ of the problem
\eqref{eq30} with $\nu=0.1$ and $\tau=0,\,3,\,6,\,7.2404$. (b) The solution plotted versus the similarity variable $\zeta/\tau^{1/4}$ for $\tau=0.001,\,0.01,\,0.1$; the dashed curve shows the similarity solution \eqref{tau0ss}.}
\end{figure}
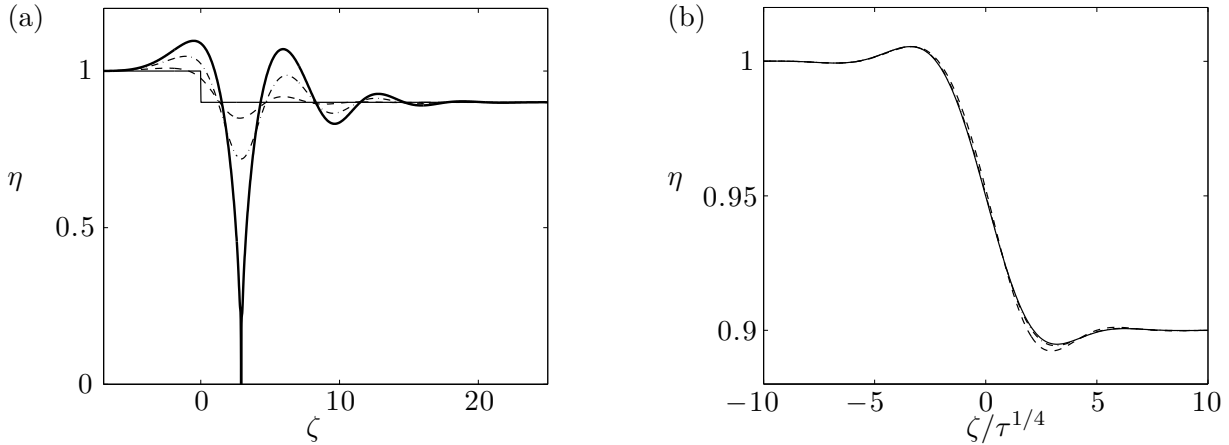

For each given value of $\nu$, we solve the problem \eqref{eq30} numerically using the method of lines. A sample solution is shown in Fig.~\ref{fig:mols1}(a), with $\nu=0.1$.
The discontinuous initial condition is instantaneously smoothed out, 
as illustrated in Fig.~\ref{fig:mols1}(b), and it can be verified that $\eta(\zeta,\tau)$ approaches the similarity solution
\begin{subequations}\label{tau0ss}
\begin{align}
&\eta(\zeta,\tau)\sim 1-\nu f\left(\zeta/\tau^{1/4}\right)
\quad\text{as }\tau\rightarrow0,\\
f(z)=\frac{1}{2}
+\frac{\Gamma(5/4)z}{\pi}\,_1\mathrm{F}_3&\left(
\frac{1}{4};\frac{1}{2},\frac{3}{4},\frac{5}{4};\frac{z^4}{256}\right) +\frac{\Gamma(-1/4)z^3}{96\pi}\,_1\mathrm{F}_3\left(
\frac{3}{4};\frac{5}{4},\frac{3}{2},\frac{7}{4};\frac{z^4}{256}
\right),
\end{align}
\end{subequations}
where $\Gamma$ and $_1\mathrm{F}_3$ denote, respectively, the gamma function and the generalized hypergeometric function.
This solution corresponds to related problems of capillary leveling of a thin film \cite{filmLeveling}; an analogous similarity solution arises in studies of the deformation of an elastica in a viscous fluid \cite{stoneduprat}.
In Fig.~\ref{fig:mols1}(a), as $\tau$ increases, the instability starts to take effect in $\zeta>0$, where the disjoining pressure term is present.
We observe wave-like disturbances that grow in amplitude, culminating in rupture of the film after a finite time $\tau_\text{rup}\approx7.24$, computed numerically. This rupture time can be determined analytically in the limit of $\nu\to 0$, and we perform this calculation next.
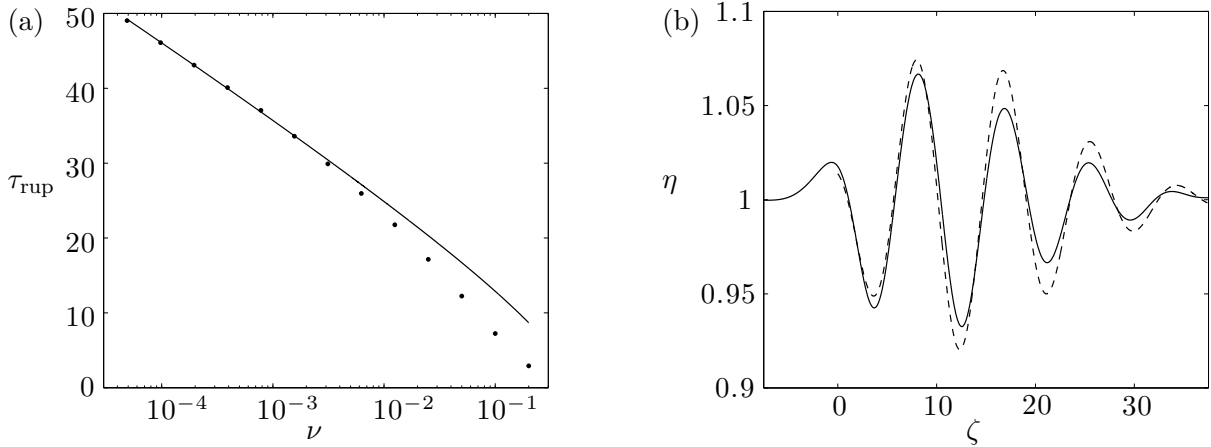
\begin{figure} [t]
\centering
\begin{minipage}[c]{1.0\linewidth}
\includegraphics[width=3in]{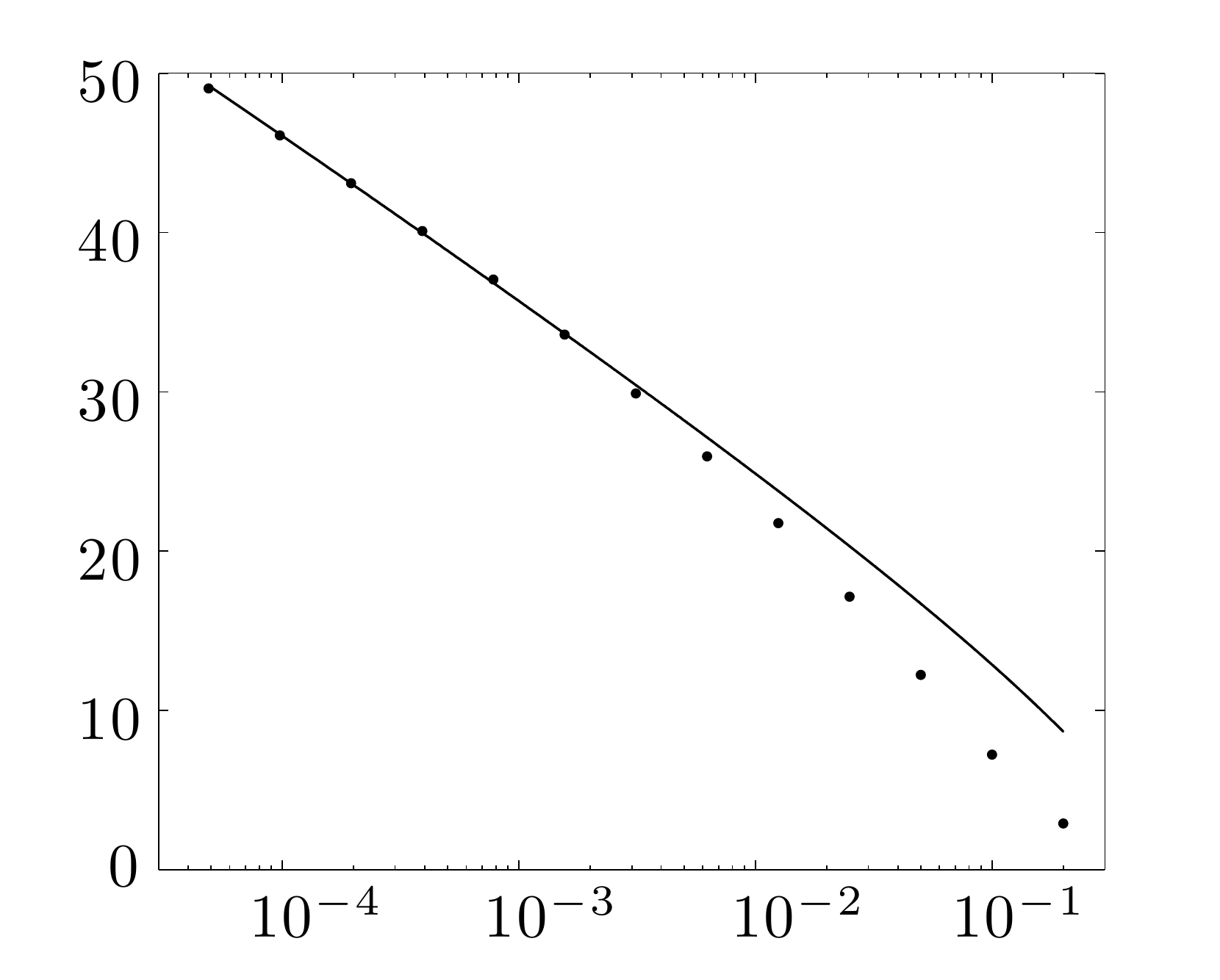}
\put(-225,95){{$\tau_\text{rup}$}}
\put(-112,0){{$\nu$}}
\put(-225,155){(a)}
\quad\quad\quad\includegraphics[width=3in]{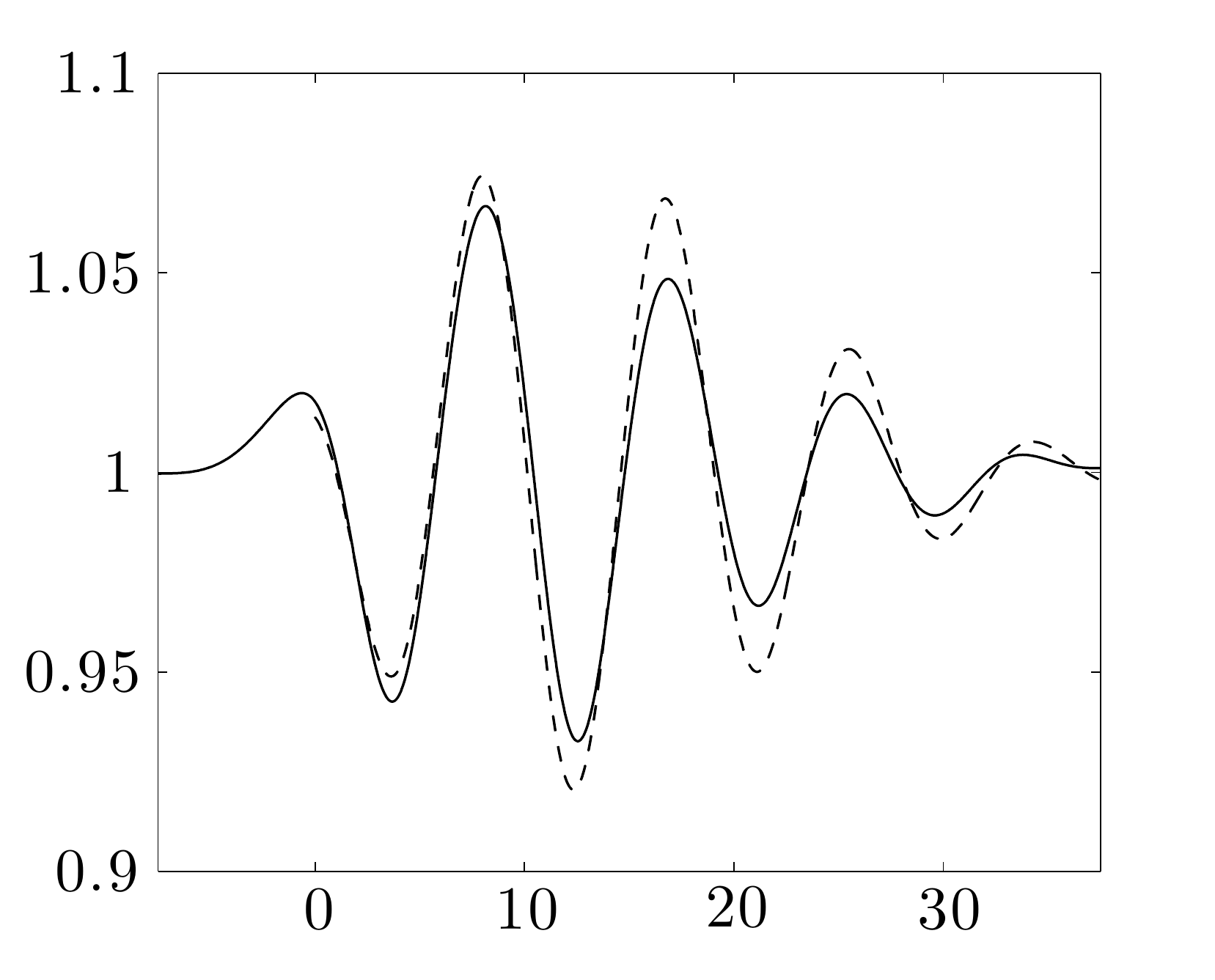}
\put(-227,95){{$\eta$}}
\put(-112,0){{$\zeta$}}
\put(-227,155){(b)}
\end{minipage}
\caption{\label{fig:tcvsnu} (a) Normalized rupture time $\tau_\text{rup}$ versus scaled van der Waals parameter $\nu$. The dots show the results obtained from numerical solutions of \eqref{eq30}; the solid curve shows the asymptotic prediction \eqref{tcasym} with $C=0.3$. (b) Numerical solution of \eqref{eq30} with $\nu=0.0001$ and $\tau=40$; the dashed curve shows the large-$\tau$ asymptotic solution $1-\nu\eta_1$, with $\eta_1$ given by Eq.~\eqref{aseta1}.}
\end{figure}

\subsubsection{Asymptotic Rupture Time} \label{ss:asympTrup}

As shown by the numerical results in Fig.~\ref{fig:tcvsnu}(a), the rupture time $\tau_\text{rup}$ increases as the value of $\nu$ decreases. To examine the limiting behavior as $\nu\rightarrow0$, we set $\eta = 1 - \nu\eta_1 + O(\nu^2)$ into \eqref{eq30}, and we find that $\eta_1$ satisfies the linear PDE
\begin{subequations}
\begin{equation}\label{ndEqtw1}
\partial_\tau\eta_1 +
\partial_{\zeta\zeta\zeta\zeta}\eta_1 +
{\partial_\zeta}\left[
\mathcal{H}(\zeta)
{\partial_\zeta}\eta_1\right] = 0,
\end{equation}
along with the boundary and initial conditions
\begin{equation}\label{ndtwbcs1}
\eta_1\rightarrow 0 \text{ as $\zeta\rightarrow-\infty$}, \qquad
\eta_1\rightarrow 1 \text{ as $\zeta\rightarrow+\infty$}, \qquad
\eta_1 = \mathcal{H}(\zeta) \text{ at $\tau=0$}.	
\end{equation}
\end{subequations}
The solution $\eta_1(\zeta,\tau)$ can be found by taking a Laplace transform in $\tau$. The resulting inversion integral is unwieldy but may be analyzed in the limit of large $\tau$ by using the method of steepest descents. A lengthy calculation (given in Appendix~A in Ref. \onlinecite{thesis_NH}) leads to the following asymptotic approximation for the solution when $\zeta>0$ and $\tau\gg1$:
\begin{equation}\label{aseta1}
\eta_1(\zeta,\tau)\sim1+\frac{1}{\sqrt{\pi}\,\tau^{3/2}}\,
\exp\left(\frac{\tau}{4}-\frac{\zeta^2}{8\tau}\right)
\,\left[
(4+\zeta)\sin\left(\frac{\zeta}{\sqrt{2}}\right)
-\sqrt{2}(2+\zeta)\cos\left(\frac{\zeta}{\sqrt{2}}\right)
\right].
\end{equation}
In Fig.~\ref{fig:tcvsnu}(b) we illustrate the accuracy of this approximation by plotting the numerical solution of \eqref{eq30} for $\eta(\zeta,\tau)$ with $\nu=0.0001$ and $\tau=40$ (solid curve) along with the approximation \eqref{aseta1}.

Equation \eqref{aseta1} implies that the maximum value of $\eta_1$
occurs when $\zeta\sim2\tau^{-1/2}$ and is given approximately by
\begin{equation}\label{eta1max}
\eta_\text{1max}(\tau)\sim2\sqrt{\frac{3}{\pi\mathrm{e}}}\,
\frac{\mathrm{e}^{\tau/4}}{\tau}
\quad\text{as }\tau\rightarrow\infty.
\end{equation}
When $\tau$ is so large that the perturbation $\eta_1$ becomes of order $1/\nu$, the asymptotic ansatz $\eta \sim 1 - \nu\eta_1$ ceases to be valid, and one must resort to numerical solution of the full governing equation \eqref{ndEqtw}. However, we can anticipate that the subsequent nonlinear evolution and rupture takes place over an $O(1)$ time-scale. We can therefore invert \eqref{eta1max} to obtain an estimate for the normalized time $\tau_\text{rup}$ taken for the film to rupture, namely,
\begin{equation}\label{tcasym}
\tau_\text{rup}\sim4\log\left(1/\nu\right)+4\log\log\left(1/\nu\right) +C,
\end{equation}
where $C$ is an $O(1)$ constant. The solid curve in Fig.~\ref{fig:tcvsnu}(a) demonstrates that \eqref{tcasym} provides a very good fit for the behavior of $\tau_\text{rup}$ as $\nu\rightarrow0$, with $C\approx0.3$.



\subsubsection{Prediction of the Critical Capillary Number} \label{ss:critCa}

By reversing the nondimensionalization \eqref{smalland}, we infer from \eqref{tcasym} the corresponding dimensional rupture time, namely,
\begin{equation}
t_\text{rup}=\frac{\kappa_0^5R\,\Ca^{1/3}}{\tilde\beta^2U}\,\tau_\text{rup}
=\frac{\kappa_0^9R\,\Ca^{13/3}}{\ell^4U}\,\tau_\text{rup}.
\end{equation}
If $t_\text{rup}$ is greater than the transit time $L/U$, then the free-surface disturbance will be swept up by the rear meniscus before rupture has time to occur (as discussed in Sec.~\ref{ss:impExp}). Therefore, rupture is not expected to occur if the capillary number exceeds a critical value $\Ca_\text{crit}$, which is found by setting $t_\text{rup}=L/U$, i.e.,
\begin{equation}
\frac{L}{R}=\frac{\kappa_0^9\,\Ca_\text{crit}^{13/3}}{\ell^4}
\,\tau_\text{rup}
\approx0.15\,\frac{\Ca_\text{crit}^{13/3}}{\ell^4}\,
\log\left(\frac{\Ca_\text{crit}}{\ell}\right). \label{Ca_crit}
\end{equation}

The result in \eqref{Ca_crit} indicates that the critical capillary number beyond which rupture is suppressed increases with increasing bubble length. This is qualitatively consistent with the experimental observations of Chen \emph{et al.} \cite{chen}. However, we note that a quantitative comparison is not attempted due to the differing assumptions between our mathematical model 
and the experiments. 



\section{Summary and Discussion} \label{sec:discuss}
\noindent In this paper, we consider the unsteady motion of an inviscid bubble advancing at a constant speed in a cylindrical capillary tube, and subject to destabilizing van der Waals forces.  
Analytical expressions for the the steady-state film profile $H_s(X; \beta)$, depending on the dimensionless van der Waals parameter $\beta$ are derived in the limiting cases of large and small $\beta$. These steady states are then evolved in time, with the front meniscus of the bubble treated separately from the rear meniscus. We find that traveling waves are created near the front meniscus, and as they are advected away from it into the thin-film region, their amplitudes grow until a rupture instability occurs. However, we find that such waves are not created in the rear meniscus, indicating that that region of the bubble remains stable. Thus, we conclude that if a traveling-wave reaches the rear bubble cap before an instability has occurred, then that wave escapes through the rear meniscus, and rupture is suppressed.

We also analyze how the Bretherton steady-state (with $\beta = 0$) is modified when attractive van der Waals forces become significant through a sudden transition from wetted to nonwetted substrates. Our asymptotic analysis demonstrates that if the bubble length is sufficiently small, then rupture is not expected to occur [cf. Eq.~\eqref{Ca_crit}]; these calculations are found to be in excellent agreement with numerical simulations of the unsteady problem for small van der Waals coefficients [cf. Fig.~\ref{fig:tcvsnu}(a)]. However, as noted in Sec.~\ref{ss:critCa}, while our asymptotic and numerical models agree qualitatively with the experimental observations of Chen \emph{et al.} \cite{chen}, a quantitative comparison is difficult due to the differing assumptions between our model and the experimental measurements. 
Our mathematical model describes a very long bubble of negligible viscosity in a circular tube, where the front and rear menisci do not interact with each other. On the other hand, the experiments measure a viscous drop of small finite length in a rectangular channel, where interaction between the front and the rear is inevitable. 
Thus, we highlight the need to (i) perform an experimental investigation of a bubble of negligible viscosity in a tube over a large range of bubble lengths; (ii) 
include viscous effects in our mathematical model to study the possible ways this can affect rupture dynamics; and (iii) develop a full numerical and analytical model which allows for arbitrary bubble lengths and shapes (and thus more complex interactions between the front and rear menisci). These considerations are the subject of ongoing work. 

In the instances that rupture cannot be completely suppressed through modification of the flow velocity, there are other ways to delay its onset. For example, \cite{MatarKumar} showed that the addition of surfactants causes an increase in the time it takes for a thin film to rupture. The effect of the surfactant is accompanied by a higher capillary number and a thicker deposited film, which may in turn suppress rupture. In their study, \cite{MatarKumar} also showed that substrate flexibility has an effect on delaying rupture; given the particular importance and desirability of rupture in certain industrial applications, (e.g., drug delivery \cite{TDD}), we highlight the importance of developing a better understanding of the dynamics of bubbles or drops in nonwetting flexible tubes. Moreover, the delay of rupture due to substrate flexibility sheds light on how the underlying substrate geometry can play a role in delaying or suppressing rupture. Several studies have sought to understand the dynamics of thin-film flow on general curved surfaces \citep{schwarzweidner, jensen, howell} and, for example, the recent work of \cite{trinh} demonstrated that substrate curvature can prevent the classical Rayleigh-Taylor instability from occurring. It would be expected that similar considerations can be made for the situation of rupture instabilities.

\begin{acknowledgments}
\textbf{Acknowledgements:} The authors thank Jens Eggers and Isabelle Cantat for helpful conversations. We also gratefully acknowledge the Oxford-Princeton Collaborative Workshop Initiative for providing an opportunity for this collaborative work.
\end{acknowledgments}

\bibliographystyle{apsrev4-1}

\end{document}